\documentclass[12pt,preprint]{aastex}

\newcommand{\pd}{\partial}
\newcommand{\md}{\mbox{d}}
\newcommand{\beq}{\begin{equation}}
\newcommand{\eeq}{\end{equation}}
\newcommand{\beqn}{\begin{eqnarray}}
\newcommand{\eeqn}{\end{eqnarray}}
\newcommand{\lppr}{\stackrel{<}{\scriptstyle \sim}}
\newcommand{\gppr}{\stackrel{>}{\scriptstyle \sim}}

\shorttitle{A microscopic analysis of shear acceleration}
\shortauthors{}

\begin{document}


\title{A microscopic analysis of shear acceleration}


\author{Frank M. Rieger and Peter Duffy}
\affil{UCD School of Mathematical Sciences, University College Dublin,
       Belfield, Dublin 4, Ireland}
\email{frank.rieger@ucd.ie; peter.duffy@ucd.ie}

\begin{abstract}
A microscopic analysis of the viscous energy gain of energetic 
particles in (gradual) non-relativistic shear flows is presented. 
We extend previous work and derive the Fokker-Planck coefficients 
for the average rate of momentum change and dispersion in the 
general case of a momentum-dependent scattering time $\tau(p)
\propto p^{\alpha}$ with $\alpha \geq 0$. We show that in contrast
to diffusive shock acceleration the characteristic shear acceleration 
timescale depends inversely on the particle mean free path which 
makes the mechanism particularly attractive for high energy seed 
particles. Based on an analysis of the associated Fokker-Planck
equation we show that above the injection momentum $p_0$ power-law 
differential particle number density spectra $n(p) \propto p^{-(1+
\alpha)}$ are generated for $\alpha >0$ if radiative energy losses 
are negligible. We discuss the modifications introduced by synchrotron 
losses and determine the contribution of the accelerated particles 
to the viscosity of the background flow. Possible implications for 
the plasma composition in mildly relativistic extragalactic jet 
sources (WATs) are addressed.
\end{abstract}

\section{Introduction}
Shear flows are a natural outcome of the density and velocity gradients 
in extreme astrophysical environments, and in the case of active 
galactic nuclei (AGNs), for example, observationally well-established 
\citep[cf.][]{lai02,lai06,rie04}. The acceleration of energetic particles 
occurring in such flows can represent an efficient mechanism for 
converting a significant part of the bulk kinetic energy of the flow 
into nonthermal particles and radiation as has been successfully shown 
in a number of contributions \citep[e.g.,][]{ost90,ost98,sta02,rie04,
rie05c}. Early theoretical progress in the field has been achieved 
based on a detailed analysis of the Boltzmann transport equation 
\citep[][]{berk81,ear88,web89}.\\
Somewhat similar to the microscopic picture for Fermi acceleration, 
shear acceleration is essentially based on the fact that particles 
can gain energy by scattering off (small-scale) magnetic field 
inhomogeneities moving with different local velocities due to being 
embedded in a collisionless shear flow with a locally changing velocity 
profile \citep[see][for a recent review]{rie05b}. In a scattering 
event particles are assumed to be randomized in direction with their 
energies being conserved in the local (comoving) flow frame. As the 
momentum of a particle travelling across a velocity shear changes 
with respect to the local flow frame, scattering leads to a net 
increase in momentum with time for an isotropic particle distribution. 
In contrast to 2nd order Fermi acceleration, however, not the random 
but the systematic velocity component of the scattering centres is 
assumed to play the important role.\\
In the present paper we present a microscopic analysis of shear 
acceleration where the underlying physical picture becomes 
most transparent. Sect.~2 gives the derivation of the Fokker-Planck 
coefficients for a simple, non-relativistic (longitudinal) gradual 
shear flow. Sect.~3 analyzes the corresponding Fokker-Planck transport 
equation providing time-dependent and steady-state solutions for 
specific cases. The contribution of the accelerated particle to the 
viscosity of the background flow is determined in Sect.~4, while
implications for mildly relativistic jet sources are addressed in 
Sect.~5.

\section{Derivation of Fokker Planck coefficients}
For a gradual non-relativistic shear flow, \citet[][]{jok90} have
calculated the Fokker-Planck coefficients using a microscopic treatment
restricted to a mean scattering time $\tau_c=$ const. independent of 
momentum \citep[cf. also][]{rie05a}. However, as under realistic
astrophysical conditions the scattering off magnetic turbulence 
structures is momentum-dependent (e.g.,  gyro-, Kolmogorov- or 
Kraichnan-type), we extend this analysis to the more general case 
where the local scattering time $\tau_c$ is a power-law function of 
momentum, i.e., $\tau_c(p)=\tau_0\,p^{\alpha}$, $\alpha \geq 0$.\\  
We consider a simple non-relativistic two-dimensional continuous shear 
flow with velocity profile given by $\vec{u} = u_z(x)\,\vec{e}_z$. 
Similarly as in \cite{jok90}, we choose a spherical coordinate system 
where $\theta$ denotes the angle between the $x$-axis and the velocity 
vector $\vec{v}=(v_x,v_y,v_z)$ of the particle, and $\phi$ is the 
azimuthal angle such that $\tan \phi =v_y/v_x$. Let $m$ be the 
relativistic mass and $\vec{p}_1$ the initial momentum of a particle 
relative to its local flow frame and $\vec{p}_2$ the corresponding 
momentum after the next scattering event. As the particle travels 
across the shear, its momentum (and thus its mean free path) changes 
with respect to the local flow frame. Denoting by $\tilde{\tau}=
<\tau>$ the mean scattering time averaged over momentum magnitude, 
one obtains
\beq
  \tilde{\tau}=\tau_c + \frac{1}{2}\,\frac{\partial \tau_c(p_1)}
                  {\partial p_1}\,\Delta p
                = \tau_c\,\left(1+\frac{1}{2}\,\alpha\,\frac{\Delta p}
                  {p_1}\right)\,,
\eeq in the limit $\Delta p/p_1 = (p_2-p_1)/p_1 \ll 1$ where collisions 
are assumed to produce only a small change in the momentum of particles, 
with large changes occurring only as a consequence of many small changes. 
Within the time $\tilde{\tau}$ a particle travels a distance $\delta 
\tilde{x} = v_x\,\tilde{\tau} = (p_1/m)\,\cos\theta\,\tilde{\tau}$ in 
the $x$-direction across the shear and the flow velocity changes by an 
amount $\delta \vec{u} = \delta \tilde{u}\,\vec{e_z}$, where
\beq
  \delta \tilde{u} = \left(\frac{\partial u_z}{\partial x}\right)\,
                     \delta \tilde{x} = \delta u\,\left(1 + 
                     \frac{1}{2}\,\alpha\,\frac{\Delta p}{p_1}\right)\,,
\eeq and $\delta u := (\partial u_z/\partial x)\,(p_1/m)\,\cos\theta\,
\tau_c$. Contenting ourselves with a Galilean transformation for the 
non-relativistic flow speeds involved, the particle's momentum relative 
to the flow will thus have changed to $\vec{p}_2 = \vec{p}_1 + m\,
\delta \vec{u}$, i.e., one obtains
\beqn\label{trafo}
    p_2^2 & = & p_1^2 + 2\,m\,\delta \tilde{u}\,\,p_{1,z} + m^2\,
              (\delta \tilde{u})^2\nonumber \\
          & = & p_1^2 \left(1+2\,\frac{m\,\delta \tilde{u}}{p_1}
              \sin\theta\cos\phi +
              \frac{m^2\,\delta \tilde{u}^2}{p_1^2}\right)\,.
\eeqn As the next scattering event is assumed to preserve the 
magnitude of the particle momentum relative to the local flow frame, 
the particle magnitude will have this value in the local frame. To 
second order in $\delta u$, one thus finds
\beqn
  p_2 \simeq & p_1 & \left(1 + \frac{m}{p_1}\,\delta u\,
                     \sin\theta\,\cos\phi
               + \frac{m^2}{p_1^2}\,\delta u^2 \right. \nonumber \\
             &  & \left. - \frac{1}{2}\,\frac{m^2}{p_1^2}\,\delta u^2\,
                     (1-\alpha)\,\sin^2\theta\,\cos^2\phi\right)\,.
\eeqn By using spherical coordinates and averaging over all momentum 
directions for an isotropic particle distribution, the Fokker-Planck 
coefficients \citep[][]{cha43} describing the average rate of momentum 
change and the average rate of momentum dispersion (associated with a 
broadening of the distribution) can be determined as follows
\beqn\label{fokker}
    \left<\frac{\Delta p}{\Delta t}\right> &:= & \frac{2\,<p_2-p_1>}{\tau_c}
          = \frac{(4+\alpha)}{15}\,p\,\left(\frac{\pd u_z}{\pd x}\right)^2
            \tau_c\;\label{momentum1} \\
   \left<\frac{(\Delta p)^2}{\Delta t}\right> &:= & \frac{2\,<(p_2-p_1)^2>}
         {\tau_c} = \frac{2}{15}\,p^2 \left(\frac{\pd u_z}{\pd x}\right)^2
         \tau_c \;,\label{momentum2}
\eeqn where the index 1 has been dropped on the right hand side, revealing 
that a net increase in momentum proportional to the square of the flow 
velocity gradient occurs with time. Obviously, the stronger the shear and 
the larger the particle mean free path, the higher the possible impact of 
scattering and thus the higher the rate of momentum change. For a constant 
scattering time, i.e., for $\alpha = 0$, these expressions reduce to those 
found by \citet{jok90}, apart from the factor of two that takes the time
average properly into account. The Fokker-Planck coefficients are related 
by the equation
\beq\label{eqa}
   \left<\frac{\Delta p}{\Delta t}\right> = \frac{1}{2\,p^2}\,
         \frac{\pd}{\pd p}
         \left(p^2 \left<\frac{(\Delta p)^2}{\Delta t}\right>\right)
         = \frac{\Gamma}{p^2}\frac{\pd}{\pd p}
           \left(p^4 \tau_c\right)\,.
\eeq where $\Gamma$ on the right hand side denotes the shear flow 
coefficient \citep[cf.][]{ear88} which, for the flow profile chosen, 
is given by
\beq
  \Gamma = \frac{1}{30}\left(\frac{\partial u_i}{\partial x_j} +
           \frac{\partial u_j}{\partial x_i}\right)^2
           -\frac{2}{45}\,\frac{\partial u_i}{\partial x_i}
            \frac{\partial u_j}{\partial x_j}
         = \frac{1}{15} \left(\frac{\partial u_z}{\partial x}\right)^2\,.
\eeq Eqs.~(\ref{momentum1}) and (\ref{eqa}) indicate that for the adopted 
scaling $\tau_c \propto p^{\alpha}$ acceleration occurs as long as 
$\alpha >-4$.
Comparison with previous approaches shows that the average rate of 
momentum increase $<\Delta p/\Delta t> = (4+\alpha)\,\Gamma\,p\,\tau_c$, 
as given by eq.~(\ref{momentum1}), agrees with the non-relativistic limit 
of the full shear acceleration coefficient derived in \citet[][]{rie04} 
and the results found by \citet[][]{berk81} and \citet[][]{ear88}.\\
Eq.~(\ref{momentum1}) implies a characteristic acceleration timescale for 
particle acceleration in non-relativistic gradual shear flows 
\beq\label{tacc}
     t_{\rm acc} = \frac{p}{<\Delta p/\Delta t>} = \frac{c}{(4+\alpha)
                   \,\Gamma\,\lambda}\,,
\eeq which is inversely proportional to the particle mean free path
$\lambda \simeq c\,\tau_c$, in remarkable contrast to diffusive shock
acceleration where $t_{\rm acc} \propto \lambda$ \citep[e.g.,][]{kir01}.
As the mean free path of a particle increases with energy it follows that
higher energy particles will be accelerated more efficiently than lower 
energy particles. This characteristic behaviour makes shear acceleration 
particularly attractive for the acceleration of high energy seed particles 
and the production of UHE cosmic rays \citep[e.g.,][]{rie05c}. 
In the case of a particle mean free path proportional to the gyroradius 
$r_g$, for example, the acceleration timescale scales with the particle 
Lorentz factor in the same way as the timescales for synchrotron and 
inverse Compton losses, respectively.

\section{Particle transport and power-law formation}
The propagation of energetic charged particles in non-relativistic 
shear flows can be cast into a Fokker-Planck-type momentum diffusion 
equation (cf. Melrose~1980). Taking synchrotron losses into account 
the isotropic phase space distribution function $f(p,t)$ (averaged 
over all momentum directions) then satisfies
\beqn\label{transport}
   \frac{\pd f(p,t)}{\pd t} & = & - \frac{1}{p^2}\,\frac{\pd}{\pd p}
          \left(p^2\,\left(\left<\frac{\Delta p}{\Delta t}\right>
           + <\dot{p}>_s \right)\, f(p,t)\right) \nonumber \\
       & & + \frac{1}{2\,p^2}\frac{\pd^2}{\pd p^2}
            \left(p^2\,\left<\frac{(\Delta p)^2}{\Delta t}\right>
             f(p,t)\right) + \tilde{Q}(p,t)\,,
\eeqn where $\tilde{Q}(p,t)$ denotes the source term, $<\dot{p}>_s$
is the synchrotron loss term given by
\beq
     <\dot{p}>_s = -\beta_s\,p^2\,,
\eeq with $\beta_s = 4\,B^2\,e^4/(9\,m^4\,c^6)$ when expressed in 
Gaussian units, and where the coefficients are related by eq.~(\ref{eqa}). 
In the absence of losses and injection, eq.~(\ref{transport}) reduces 
to a momentum space diffusion equation 
\beq
{\pd f\over\pd t}= {1\over p^2}{\pd\over\pd p}
\left( p^2 D{\pd f\over\pd p}\right)
\eeq where $D=\Gamma p^{2+\alpha}\tau_0$ is the momentum space 
diffusion coefficient. Eq.~(\ref{eqa}) is known as the principle of 
detailed balance \citep[e.g.,][]{bla87} and is a condition which 
must be satisfied for the Fokker-Planck equation to reduce to the 
case of pure diffusion.\\
It is interesting to note that second order Fermi acceleration as a 
result of scattering from forward and reverse Alfven waves is also 
described by momentum space diffusion \citep[][]{ski75}. It is also 
possible to analyse such a process on the microscopic level, deriving 
the Fokker Planck coefficients and demonstrating the principle of 
detailed balance, see \citet[][]{duf05} for details.
 
In the following eq.~(\ref{transport}) will be used to analyze the 
the evolution of the particle distribution function for some 
particular cases. 

\subsection{Time-dependent solutions for impulsive source}\label{time}
\begin{figure}
\epsscale{.80}
\plotone{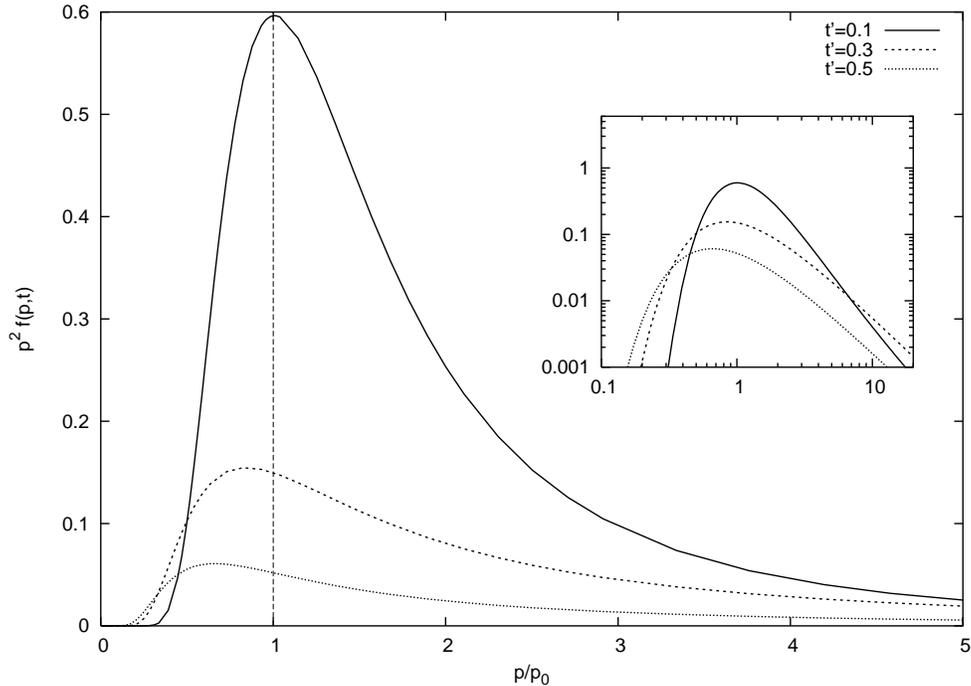}
\caption{Characteristic evolution of the differential particle number 
density $n(p,t) \propto p^2 f(p,t)$ for $\alpha=1$ (cf. 
eq.~[\ref{besselI_sol}]) as a function of momentum $p/p_0$ in the case 
of impulsive injection of particles with $p_0$ at $t=0$. 
The distribution function is plotted at three different times $t$ where 
$t=t'\cdot t_c$ with $t_c=1/(\Gamma \tau_0 p_0)$. The inlay shows the 
same in double logarithm representation and already indicates the 
formation of a power-law tail $n(p,t) \propto p^{-2}$ for $t' \geq 
0.3$. \label{besselI}}
\end{figure}

Consider first the case where radiative synchrotron losses are 
negligible and particles are injected monoenergetically at time 
$t_0=0$ with momentum $p_0$, i.e., $\tilde{Q}(p,t) = Q\,
\delta(p-p_0)\,\delta(t)$. Eq.~(\ref{transport}) then reduces to 
\beq\label{impulsive}
  \frac{\pd f(p,t)}{\pd t} = \frac{1}{p^2}\frac{\pd}{\pd p}
                             \left(\Gamma\,p^{4+\alpha}
                            \tau_0 \frac{\pd f(p,t)}{\pd p}\right) 
                            + Q\,\delta(t)\delta(p-p_0)\,.
\eeq The same equation is obtained if one uses the generalized 
Parker transport equation \citep[][eq.~6]{ear88} for the shear flow 
profile under consideration. As shown in the appendix the 
mathematical solution of eq.~(\ref{impulsive}) for $\alpha \neq 
0$ takes the form, see eq.~(\ref{kep_sol}),
\beq\label{besselI_sol}
   f(p,t) = \frac{Q  p_0^{-(\alpha+1)}}{|\alpha| \Gamma \tau_0 t} 
            \left(\frac{p_0}{p}\right)^{(3+\alpha)/2}
            \exp\left(-\frac{p^{-\alpha}+p_0^{-\alpha}}{\alpha^2 
                \Gamma \tau_0 t}\right) I_{|1+3/\alpha|}\left(
                \frac{2}{\alpha^2 \Gamma \tau_0 p_0^{\alpha} t} 
                 \left[\frac{p}{p_0}\right]^{-\alpha/2}\right)\,,
\eeq where $I_{\nu}(z)$ is the modified Bessel function of the 
first kind \citep[cf.][]{abr72}. The corresponding differential
particle number density $p^2 f$ is illustrated in 
Fig.~(\ref{besselI}) using a momentum dependence of $\alpha=1$. 
Obviously, the position of the peak shifts to the left and 
decreases with time (thus ensuring particle number conservation) 
while the distribution becomes broader and the relative strength 
of its tail increases (as a consequence of dispersion and 
acceleration). For small $z \rightarrow 0$ the modified Bessel 
function is known to approach $I_{\nu}(z) \sim (z/2)^{\nu}/
\tilde{\Gamma}(\nu+1)$ for $\nu \neq -1,-2,..$ with 
$\tilde{\Gamma}(\nu+1)$ denoting the Gamma function. It thus
follows that for $\alpha >0$ and $p\gppr p_0$ the phase space 
distribution approaches a power-law $f \propto p^{-(3+\alpha)}$ 
on a characteristic timescale $t_c =1/(\alpha^2 \Gamma \tau_0 
p_0^{\alpha})$ \citep[cf.][]{ber82}. For $p\geq p_0$ the 
time-integrated solution of eq.~(\ref{impulsive}) will thus 
show the same power-law behaviour, cf. eq.~(\ref{powerlaw}).\\
Note that the Fokker Planck description employed is only 
valid at late and not at very early stages since the 
resultant transport equation is essentially non-causal, 
i.e., eq.~(\ref{transport}) is a parabolic partial 
differential equation. Its solutions thus share with those 
of the heat equation the mathematical properties of an 
infinite speed of propagation (i.e., the distribution is 
non-zero everywhere for $t>0$), smoothing of singularities 
and decrease and broadening with time. In general, the 
problem of infinite propagation may be overcome by using
a modified diffusion equation of the (hyperbolic) 
telegrapher's type \citep[e.g.,][]{cat48,mor53}. 
Formally, eq.~(\ref{transport}) relies on the diffusion 
approximation and therefore only provides an adequate 
description of the particle transport for times much larger
than the mean scattering time, i.e., as long as $\tau(p) 
\ll t$ which constrains the maximum momentum range for a 
given time $t$ over which the solution eq.~(\ref{besselI_sol}) 
may be considered appropriate. Complementary, from a 
physical point of view particle energization is expected 
to occur only on a characteristic timescale determined by 
the systematic acceleration in eq.~(\ref{tacc})\citep[e.g.,]
[]{bal92}, which constrains the characteristic maximum 
momentum $p(t)$ achievable within a given time interval.

\subsection{Time-integrated (steady-state) solutions} 
\begin{figure}
\epsscale{.80}
\plotone{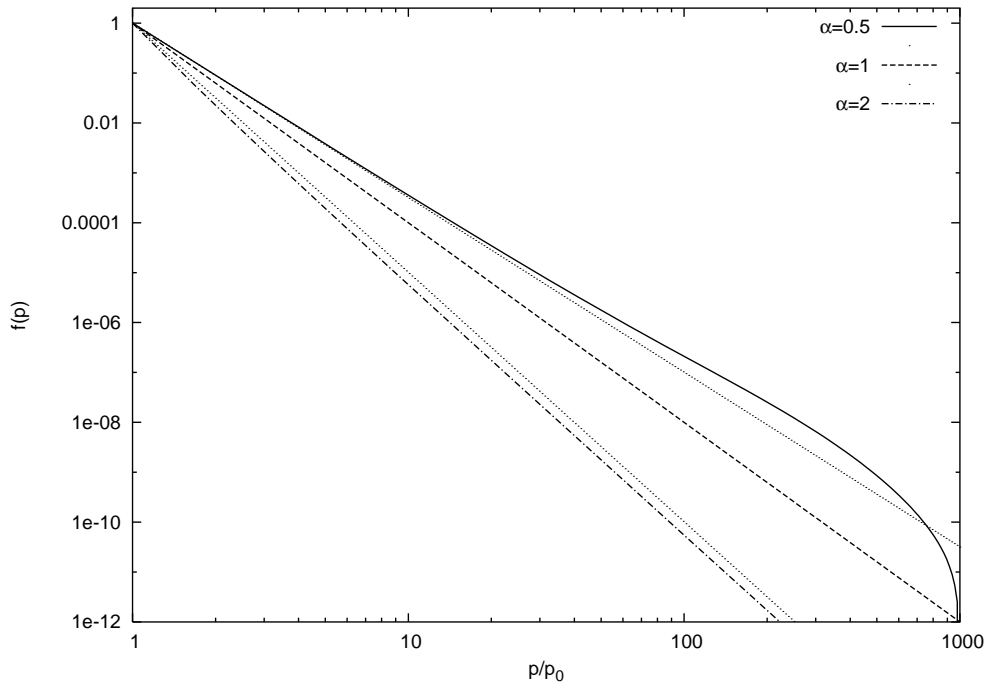}
\caption{Evolution of the normalized phase space particle
distribution function $f(p)$ as a function of momentum $p/p_0$ 
for different power indices of the mean scattering time $\tau 
\propto p^{\alpha}$. The thin-dotted lines are drawn to guide 
the eyes and correspond to power-law distributions $f(p) 
\propto p^{-3.5}$ and $p^{-5}$, respectively. For $\alpha=0.5$ 
the maximum particle  momentum, at which acceleration is balanced 
by losses, has been chosen to be $p_{\rm max}=1000\,p_0$, whereas
for $\alpha=2$ a minimum momentum $p_{\rm min}=p_0/2$ has been 
used.\label{distr}}
\end{figure}

Defining the time-integrated phase space distribution function
\beq
   F(p) = \int_0^{\infty} f(p,t) \mathrm{dt}
\eeq and integrating the Fokker Planck equation~(\ref{transport}) 
over time assuming a monoenergetic source term $\tilde{Q}(p,t)= 
Q\,\delta(p-p_0)\,\delta(t)$ gives
\beq\label{steady}
  \frac{\pd}{\pd p}\left(\Gamma p^{4+\alpha} \tau_0 
  \frac{\pd F(p)}{\pd p}\right) + \frac{\pd}{\pd p}
  \left(\beta_s p^4 F(p)\right) = -Q p_0^2\delta(p-p_0)\,.    
\eeq Note that the time-integrated function $F(p)$ is proportional 
to the steady state solution of eq.~(\ref{transport}) for 
continuous injection since eq.~(\ref{steady}) coincides with 
eq.~(\ref{transport}) if $\pd f/\pd t = 0$. The general solution 
of eq.~(\ref{steady}) is of the form
\beq
    F(p) = \frac{Q  p_0^2}{\Gamma \tau_0} \left[c_1-H(p-p_0)\right] 
             e^{-\chi(p)} \int_{p_0}^{p} \mathrm{dp'}
             \frac{e^{\chi(p')}}{p'^{4+\alpha}}
             + c_2 e^{-\chi(p)}
\eeq where $H(p)$ is the Heaviside step function, $c_1,c_2$ are 
constants to be determined by the boundary conditions, and
\beqn
  \chi(p) = \left\{ \begin{array}{rl}
           - \frac{\beta_s}{(\alpha-1)\,\Gamma\,\tau_0}\,p^{-(\alpha-1)}
           & \;{\rm for}\;\; \alpha \neq 1 \\
           \frac{\beta_s}{\Gamma\,\tau_0}\,\ln p
           & \;{\rm for}\;\; \alpha = 1
          \end{array}\,. \right.
\eeqn If synchrotron losses are negligible ($\beta_s = 0$) the 
time-integrated phase space distribution follows a simple power law 
above $p_0$ with momentum index $-(3+\alpha)$ \citep[cf.][]{rie05a}, 
i.e., one finds
\beq\label{powerlaw}
     F(p) = \frac{Q p_0^2}{(3+\alpha) \Gamma \tau_0}
            \left(\frac{1}{p^{3+\alpha}} H(p-p_0) 
             + \frac{1}{p_0^{3+\alpha}} H(p_0-p)\right)\,.
\eeq for $\alpha >0$ and boundary conditions $F(p\rightarrow \infty) 
\rightarrow 0$ and $\pd F/\pd p \rightarrow 0$ for $p\rightarrow 0$. 
Thus, in the case of a mean scattering time scaling with the gyroradius 
(i.e., Bohm case: $\alpha =1$), for example, this corresponds to a 
differential power law particle number density $n(p) \propto p^2 F(p) 
\propto p^{-2}$ above $p_0$ and $n(p) \propto p^2$ below $p_0$.\\
If losses are non-negligible, the solutions become more complex. 
For $\alpha <1$ we end up with expressions involving exponential 
integrals \citep[cf.][]{abr72}. The characteristic evolution of the 
phase space particle distribution function is illustrated in 
Fig.~(\ref{distr}). For $\alpha<1$ the acceleration efficiency is 
generally constrained by synchrotron losses, which results in a 
cut-off at $p_{\rm max}=([4+\alpha]\,\Gamma\,\tau_0/\beta_s)^{(1/
[1-\alpha])}$ where the acceleration timescale $t_{\rm acc}$ (cf. 
eq.~[\ref{tacc}]) matches the cooling timescale $t_{\rm syn} = 1/
(\beta_s\,p)$. If $p \ll p_{\rm max}$ cooling effects are negligible 
and the particle distribution follows a power-law as described in
eq.~(\ref{powerlaw}). For $\alpha=1$ the ratio $t_{\rm acc}/t_{\rm 
syn}$ becomes independent of momentum. Hence if conditions are such 
that $\beta_s/(\Gamma\,\tau_0)<5$ (cf. eq.~[\ref{transport}]) is 
satisfied, particle acceleration is no longer constrained by 
synchrotron losses and for $\beta_s/(\Gamma\,\tau_0)<4$ the resultant 
particle distribution above $p_0$ becomes $F(p)=Q p_0^2/(4\,\Gamma\,
\tau_0-\beta_s)\,p^{-4}$ as expected from eq.~(\ref{powerlaw}). 
For $\alpha >1$ shear acceleration becomes possible (and is essentially 
unconstrained by synchrotron losses) when particles are injected with 
momenta above a threshold $p_{\rm min}=(\beta_s/([4+\alpha]\,\Gamma\,
\tau_0))^{1/[\alpha-1]}$. For $p \gg p_{\rm min}$ cooling effects are 
completely negligible and the particle distribution approaches a 
power-law with momentum index as given in eq.~(\ref{powerlaw}).\\
Note that if losses are unimportant a proper physical steady state 
situation may be achieved in instances where particles are continuously 
injected with momentum $p_0$ and considered to escape above a fixed 
momentum $p_{\rm max} \gg p_0$ where the associated particle mean free 
path becomes of order of the size of the system. 
Formally, this may be done by adding a simple momentum-dependent 
escape term $Q (p_0/p_{\rm max})^2 \delta(p-p_{\rm max})$ on the 
right hand side of eq.~(\ref{steady}).

\section{Cosmic-ray viscosity}
As particles gain energy by scattering off inhomogeneities embedded 
in a background flow, efficient shear acceleration essentially 
draws on the kinetic energy of that flow. In principle, efficient 
cosmic ray acceleration may thus cause a non-negligible deceleration 
even for (quasi-collimated) large-scale relativistic jets, although 
a proper assessment of its significance in comparison with other 
mechanisms (e.g., entrainment) will require detailed source-specific
modelling. In any case, for the purposes of our (non-relativistic) 
analysis here, the resultant dynamical effects on the flow can be 
modelled by means of an induced viscosity coefficient $\eta_s > 0$ 
that describes the associated decrease in flow mechanical energy per 
unit time, e.g.,
\begin{equation}
\dot{E}_{\rm kin} =-\eta_s
\int \left(\frac{\partial u_z}{\partial x}\right)^2
     \mbox{\rm d}V\,,
\end{equation} for our case of a (non-relativistic) two-dimensional 
gradual shear flow $u_z(x)\,\vec{e}_z$ \citep[][\S 16]{lan82}. 
We may determine this viscosity coefficient using $\dot{E}_{\rm kin} 
= -\dot{E}_{\rm cr}$, where $E_{\rm cr}=\int \epsilon_{\rm cr} 
\mbox{\rm d}V$ is the energy gained by the cosmic ray particles 
\citep[cf. also][]{ear88}. If particle injection is described by
a continuous source term $Q \delta(p-p_0)$, the density 
$\dot{\epsilon}_{\rm cr}$ of power gained becomes
\begin{equation}\label{viscos}
\dot{\epsilon}_{\rm cr} = 4\pi\int_0^{\infty}
     p^2\,E(p)\,\frac{\partial f(p,t)}{\partial t}
     \,\mbox{\rm d}p - 4\,\pi\,Q\,E(p_0)\,p_0^2\,,
\end{equation} where $E(p)$ denotes the relativistic kinetic energy. 
Note that due to the constraints discussed at the end of \S~\ref{time},
special care has to be taken if one wishes to evaluate the integral 
in eq.~(\ref{viscos}), as done in \citet[][]{ear88}, by replacing 
the time-derivative of $f$ with the Fokker-Planck expression of 
eq.~(\ref{impulsive}) for the chosen source term.\footnote{Note 
that as shown above, we cannot choose independently a power-law 
index for $f$ \citep[e.g., as suggested in][]{ear88} once we have 
chosen a momentum index for the particle mean free path.}
From an astrophysical point of view, we are most interested in 
steady-state type situations where particles are injected 
quasi-continuously with momentum $p_0$ and, in the absence of 
significant radiative losses, considered to escape above a momentum 
threshold $p_{\rm max}$ at which $\lambda(p_{\rm max})$ becomes 
larger than the width of the acceleration region. The density of power 
gained then becomes $\dot{\epsilon}_{\rm cr} \simeq 4 \pi Q p_0^3 c 
\left(p_{\rm max}/p_0 -1 \right) \simeq 4 \pi Q p_0^2\,p_{\rm max}\,c$ 
for $p_{\rm max} \gg p_0 \gg m_0\,c^2$. This implies a viscosity 
coefficient 
\begin{equation}\label{visc_coeff}
  \eta_s \simeq \frac{3\,\alpha}{15}\,\lambda(p_0)\,
         n_0\,p_{\rm max}\,,
\end{equation} for $\alpha >0$, where $\lambda(p_0) \simeq \tau_c(p_0) 
c$ denotes the mean free path for a particle with momentum $p_0$, and 
$n_0 = 4\pi \int_0^{p_{\rm max}}p^2 f(p) \md p$ is the number density
of cosmic-ray particles in the acceleration region.\\

\section{Applications}
Efficient shear acceleration of cosmic-ray particles is likely to occur 
in a number of powerful jet sources including galactic microquasars and 
extragalactic FR~I and FR~II sources \citep[e.g.,][]{sta02,lai02,lai06}. 
Whereas a proper analysis of powerful AGN-type jets requires a fully
relativistic treatment \citep[cf.][]{rie04}, application of the results 
derived above may allow useful insights in the case of moderately 
relativistic jet sources. As an example, let us consider the possible 
role of shear acceleration in wide angle tailed radio galaxies (WATs), 
deferring a detailed discussion of particle acceleration in microquasar 
jets to a subsequent paper. WATs are central cluster galaxies and appear 
as "hybrid" sources \citep[][]{jet06} showing both FR~I and FR~II 
morphologies. Their inner jets extend tens of kiloparsec, seem to 
have central speeds in the range $v_j \simeq (0.3-0.7)$ c (provided 
Doppler hiding effects can be neglected), are apparently very well 
collimated on the kiloparsec scale and exhibit spectra close to $S(\nu) 
\propto \nu^{-0.5}$, with evidence pointing to a steeper spectrum sheath 
likely to be induced by the strong interactions between the jet and its 
environment \citep[e.g.,][]{kat99,har99,har05,jet06}. Phenomenological 
studies suggest a velocity transition layer width for the large-scale 
jet of $\Delta r = \xi$ kpc ($\xi \lppr 0.5$) and a fiducial jet 
magnetic field strength of $B = 10^{-5} b_0$ Gauss ($b_0 \gppr 1$ 
for equipartition).
As a likely scenario let us consider the case where energetic seed 
particles required for efficient shear acceleration are provided by 
internal shock-type (Fermi I) processes operating in the jet 
interior and giving rise to a power-law proton distribution $f_Q(p) 
= n_Q p_1/(4 \pi p^4)$ in the momentum range $p_1 \simeq 2 m_p c 
\leq p \leq p_2$, where $p_2 \gg p_1$ may be determined either from 
the condition of lateral confinement (e.g., $p_2 \sim 10^{10} \xi 
b_0 m_p c$ for $\lambda \sim r_g$) or via the balance of radiative 
synchrotron losses, and where $n_Q$ is the number density of 
accelerated particles. The minimum (Bohm diffusion) timescale for 
non-relativistic shock acceleration is of the order $t_{\rm acc}(p) 
\sim 6 r_g\,c/u_s^2$ where $u_s$ is the shock speed as measured in 
the upstream frame \cite[e.g.,][]{rie05c}. For typical parameters 
$u_s \sim 0.1$c, $\lambda \sim r_g$ and a simple linear shear profile 
with $\Gamma \sim (v_j/\Delta r)^2/15$, shear acceleration (cf. 
Eq.~[\ref{tacc}]) will dominate over shock-type processes for protons 
with Lorentz factors above $\gamma_c \sim 10^9 b_0 \xi$, so that 
protons with momenta $p \geq p_c=\gamma_c m_p c$ may be considered 
as being effectively injected into the shear acceleration mechanism, 
resulting in a rate of injected particles of $Q \simeq n_s/t_0$, 
where $n_s \simeq n_Q (p_1/p_c)\sim 2 \cdot 10^{-9} n_Q$, $t_0 \sim 
t_{\rm acc}(p_c)/P_{\rm esc}$ and $P_{\rm esc} \simeq 4 u_2/c$ is 
the escape probability of a particle from the shock (with $u_2$ 
measured in the shock frame). This yields a viscosity coefficient 
$\eta_s \sim n_s p_{\rm max} c/(15 \Gamma t_0)$. 
Now, the Navier-Stokes equation imply that in a viscous flow of density 
$\rho_f$ and viscosity coefficient $\eta_s$, the characteristic viscous
damping timescale $T$ for the decay of velocity structures of size $L$ 
is of the order $T \sim \rho_f L^2/\eta_s$ \citep[e.g.,][]{ear88}. For 
$L \simeq \Delta r$ the numbers estimated above give a characteristic 
decay timescale $T$ of the order 
\beq
  T \sim  \left(\frac{\rho_f}{\gamma_{\rm max} n_s m_p}\right) 
          \left(\frac{v_j}{c}\right)^2 t_0 
          \sim 2 \cdot 10^4 \left(\frac{\rho_f}{n_Q m_p}\right)
          \left(\frac{\xi}{0.3}\right) \mathrm{yr}\,.
\eeq Requiring $T$ to be larger than the timescale $t_l$ set by the 
apparent stability of the jet flow, i.e., $t_l \sim L_j/v_j \sim 2 
\cdot 10^5$ yr for a typical jet length $L_j \sim 30$ kpc and $v_j 
\sim 0.5$ c, gives a minimum density ratio contrast of cold matter 
to energetic protons $(\rho_f/[n_Q m_p]) \sim 10\,(0.3/\xi)$. 
Therefore, if similar to the supernova remnant or solar wind case 
atmost a fraction $\sim 0.01$ of the overtaken thermal particles 
is injected into the shock acceleration process \citep[e.g., see]
[]{tra91,duf95,bar99}, significant velocity decay effects 
are unlikely even if the total power of WAT jets should be 
still dynamically dominated by a cold (thermal) proton component. 
We note that such a case (slight thermal proton dominance) appears 
not impossible and may indeed be consistent with the notion that 
WATs are intermediate stages in a broader framework where FR~I 
appear as pair-dominated sources and FR~II jets as mainly composed 
of electrons and protons \citep[e.g.,][]{cel97}, although 
circumstantial evidence based on jet bending seems to suggest that
WAT jets are rather light \citep[][]{har05,jet06}.

\section{Conclusions}
Turbulent shear flows are widely expected in astrophysical environments. 
Using a microscopic analysis, we have shown that, in the absence of strong
synchrotron losses, the acceleration of energetic particles occurring in 
such flows can give rise to power-law differential particle number densities 
$n(p) \propto p^{-(1+\alpha)}$ above the injection momentum $p_0$ for a 
scattering time $\tau_c \propto p^{\alpha}, \alpha >0$. Dependent on the
details of the underlying turbulence spectrum, shear acceleration may thus
allow for different power-law indices. As efficient shear acceleration 
generally requires sufficiently energetic seed particles, this implies an 
interesting corollary: if energetic seed particles are provided by shock-type 
acceleration processes, the take over by shear acceleration may reveal 
itself by a change of the power-law index above the corresponding energy 
threshold. Perhaps even more interesting is the fact that the characteristic
timescale for particle acceleration in gradual shear flows is inversely 
proportional to the particle mean free path, i.e., $t_{\rm acc} \propto
1/\lambda$. Shear acceleration thus leads to a preferred acceleration of
particles with higher magnetic rigidity. Indeed, detailed analyses show
that in realistic astrophysical circumstances efficient shear acceleration 
works usually quite well for protons, but appears restricted for electrons 
\citep[e.g.,][]{rie04,rie05c}. As shear acceleration essentially draws on 
the kinetic energy reservoir of the background flow, the associated viscous 
drag force can, depending on the intrinsic plasma characteristics, 
significantly contribute to a deceleration of the large-scale jet flow. 
This suggests that shear acceleration may have important implications for 
our understanding of the acceleration of cosmic ray particles, the plasma 
composition and the velocity evolution in astrophysical jet sources.

\acknowledgments
Financial support through a Marie-Curie and a Cosmogrid Fellowship
and discussions with John Kirk are gratefully acknowledged.  
Useful comments by the anonymous referee that strengthen the 
application are appreciated.

\appendix
\section{Derivation of time-dependent solutions for impulsive sources}
For an impulsive source where $Q$ particles are assumed to be 
injected with momentum $p_0$ at time $t=0$, the shear Fokker-Planck 
equation reads
\beq\label{impulsive_app}
  \frac{\pd f(p,t)}{\pd t} = \frac{1}{p^2}\frac{\pd}{\pd p}
                             \left(\Gamma\,p^{4+\alpha}
                            \tau_0 \frac{\pd f(p,t)}{\pd p}\right) 
                            + Q\,\delta(t)\delta(p-p_0)\,.
\eeq For $\alpha \neq 0$ we may choose new variables $\hat{t}:=
\Gamma \tau_0 t$ and $x:=p^{-\alpha}$, for which the homogeneous
part of eq.~(\ref{impulsive_app}) becomes  
\beq\label{impulsive_trafo}
  x \frac{\pd^2 f(x,\hat{t})}{\pd x^2} 
  -\left(\frac{3}{\alpha}\right)\frac{\pd f(x,\hat{t})}{\pd x} 
  - \frac{1}{\alpha^2} \frac{\pd f(x,\hat{t})}{\pd \hat{t}} =0\,.
\eeq Eq.~(\ref{impulsive_trafo}) can be identified with the 
Kepinski partial differential equation \citep[][]{kep06}
\beq\label{kepinski}
\frac{\pd^2 f}{\pd x^2} + \frac{m+1}{x} \frac{\pd f}{\pd x}
    - \frac{n}{x} \frac{\pd f}{\pd \hat{t}} =0\,, 
\eeq for $n \equiv 1/\alpha^2$ and $m \equiv -(3+\alpha)/
\alpha$. For the initial condition $f(\hat{t}=0,x)=
\tilde{f}(x)$ the solution of the Kepinski partial differential 
equation is known and of the form \citep[][]{kep06}
\beq
 f(x,\hat{t})= \frac{n}{\hat{t}} \int_0^\infty 
             \left(\frac{\lambda}{x}\right)^{m/2} 
             \exp\left(-n \frac{x+\lambda}{\hat{t}}\right) 
             I_{|m|}\left(2 n \frac{\sqrt{x \lambda}}{\hat{t}}\right)
             \tilde{f}(\lambda) \md  \lambda\,,
\eeq where $I_{\nu}(z)$ denotes the modified Bessel function of 
the first kind \citep[cf.][]{abr72}. For our initial condition 
$Q \delta(p-p_0)$ at $t=0$ we have
\beq 
  \tilde{f}(\lambda) = Q |\alpha| p_0^{-(\alpha+1)} 
              \delta(\lambda-\lambda_0)\,.
\eeq where $\lambda_0=p_0^{-\alpha}$. Using the original set of
variables $(p,t)$, the full solution of eq.~(\ref{impulsive_app}) 
for $\alpha \neq 0$ thus becomes
\beq\label{kep_sol}
   f(p,t) = \frac{Q  p_0^{-(\alpha+1)}}{|\alpha| \Gamma \tau_0 t} 
            \left(\frac{p_0}{p}\right)^{(3+\alpha)/2}
            \exp\left(-\frac{p^{-\alpha}+p_0^{-\alpha}}{\alpha^2 
                \Gamma \tau_0 t}\right) I_{|1+3/\alpha|}\left(
                \frac{2}{\alpha^2 \Gamma \tau_0 p_0^{\alpha} t} 
                 \left[\frac{p}{p_0}\right]^{-\alpha/2}\right)\,.
\eeq Eq.~(\ref{kep_sol}) agrees with the solution presented in
\citet[][]{ber82} and can be shown to reduce to $f(p,t) \rightarrow 
Q \delta(p-p_0)$ in the limit $t\rightarrow 0^+$. In the case of 
$\alpha=-2$ we may use that   
\beq
 I_{\frac{1}{2}}(z)=\frac{1}{\sqrt{2 \pi z}} 
                     \left(e^z - e^{-z}\right) 
\eeq to obtain 
\beq
  f(p,t) = \frac{Q p_0}{p \sqrt{4 \pi \Gamma \tau_0 t}}
      \left(\exp\left[-\frac{(p-p_0)^2}{4 \Gamma \tau_0 t}\right]
       - \exp\left[-\frac{(p+p_0)^2}{4 \Gamma \tau_0 t}
       \right]\right)\,,
\eeq which agrees with the solution derived for this particular 
case by \citet[][]{ear88}, see their eq.~(13).\\ 
For $\alpha =0$ we may use the variables $z:=\ln p +3 \hat{t}$, 
where $\hat{t}:=\Gamma \tau_0 t$, to obtain the characteristic 
diffusion equation $\pd^2 f/\pd z^2 =\pd f/\pd \hat{t}$ for the 
homogeneous part of eq.~(\ref{impulsive_app}) with fundamental
solution $f(z,\hat{t}) =\exp(-z^2/[4 \hat{t}])/(4 \pi \hat{t})^{1/2}$.
Thus, for $\alpha=0$ the solution of eq.~(\ref{impulsive_app})
becomes \citep[cf. also][]{kar62}
\beq
 f(p,t) = \frac{Q}{p_0 \sqrt{4 \pi \Gamma \tau_0 t}}
          \exp\left(-\left[\ln \frac{p}{p_0}+ 3 \Gamma \tau_0 t
                     \right]^2/(4 \Gamma \tau_0 t)\right)\,.
\eeq

\clearpage

\end{document}